\documentclass[%
 reprint,
 amsmath,amssymb,
prb,
]{revtex4-1}

\usepackage{graphicx}
\usepackage{dcolumn}
\usepackage{bm}

\begin{document}

\preprint{APS/123-QED}

\title{First-principles study of carbon impurities in CuIn$_{1-x}$Ga$_x$Se$_{2}$, present in nonvacuum synthesis methods}

\author{J. Bekaert}
\email{Jonas.Bekaert@uantwerpen.be}
\affiliation{%
 CMT-group and EMAT\\
 Department of Physics, University of Antwerp\\
 Groenenborgerlaan 171, B-2020 Antwerp, Belgium
}%
\author{R. Saniz}%
\author{B. Partoens}
\author{D. Lamoen}
 
\affiliation{%
 CMT-group and EMAT\\
 Department of Physics, University of Antwerp\\
 Groenenborgerlaan 171, B-2020 Antwerp, Belgium
}%

\date{\today}

\begin{abstract}
A first-principles study of the structural and electronic properies of carbon impurities in CuIn$_{1-x}$Ga$_x$Se$_{2}$ is presented. Carbon is present in organic molecules in the precursor solutions used in nonvacuum growth methods, making more efficient use of material, time and energy than traditional vacuum methods. The formation energies of several carbon impurities are calculated using the hybrid HSE06 functional. C$_{\mathrm{Cu}}$ acts as a shallow donor, C$_{\mathrm{In}}$ and interstitial C yield deep donor levels in CuInSe$_{2}$, while in CuGaSe$_{2}$ C$_{\mathrm{Ga}}$ and interstitial C act as deep amphoteric defects. So, if present, these defects reduce the majority carrier (hole) concentration by compensating the acceptor levels and become trap states for the photogenerated minority carriers (electrons). However, the formation energies of the calculated carbon impurities are high, even under C-rich growth conditions. Therefore, these impurities are not likely to form and will probably be expelled to the intergranular region and out of the absorber layer. 
\end{abstract}

\maketitle


\section{Introduction}

Currently the most common synthesis methods for polycrystalline CuIn$_{1-x}$Ga$_x$Se$_{2}$ (CIGS) layers (with $0 \leq x \leq 1$) can be divided into two categories: coevaporation and sequential deposition methods. The most efficient CIGS thin-film photovoltaic cells, with efficiencies exceeding 20\% both on glass substrates \cite{Jackson2} and flexible substrates \cite{Chirila}, are made through coevaporation of Cu, In, Ga and Se in a vacuum chamber, using a three-stage process \cite{Kodigala,Singh}. On the other hand, in the sequential deposition methods first a precursor material is prepared. The precursor is subsequently deposited on a substrate and annealed, inducing the chalcogenization reaction. The main advantage of the sequential methods over coevaporation is that large-area films can be grown \cite{Singh}. The standard sequential method consists of preparing a precursor of Cu, In and Ga by sputtering or thermal evaporation, both vacuum-based. In the next stage, called selenization, the precursor is exposed to Se while at the same time it is annealed at $\sim 400-500~^{\circ}$C. Coevaporation and the preparation of the precursor for the standard sequential method both involve vacuum conditions and are therefore afflicted with several problems. First of all, material losses of 20 to 50 \% are common \cite{Kodigala}. The main cause is unintentional deposition of material on the vacuum chamber walls. Also, creating and maintaining a vacuum demands a high energy input. Furthermore, vacuum-based methods tend to be relatively slow. To overcome these limitations, nonvacuum-based sequential synthesis methods are gaining interest. These methods have been shown to reduce the material losses to almost zero \cite{Tiwari2009}. They are also known as wet methods, as the precursor is usually included in a solution. Based on differences in the precursor material and deposition method one distinguishes (i) coating through electrochemical reactions in a solution, (ii) coating with a molecular precursor solution by mechanical means and (iii) particulate-based processes \cite{Tiwari2009}. In the latter case, the particulates are solid nanoparticles, usually consisting of multiple oxide or selenide phases of Cu, In and Ga, dispersed in an organic solvent. In this way, they form a sort of ink that is coated onto the substrate by printing, spraying or spin coating. The organic solvents that have been reported in the scientific literature include a mixture of methanol and pyridine \cite{Schulz1998}, a mixture of ethanol, terpineol and ethyl cellulose \cite{Lee2011} and 1,5-pentanediol \cite{Zaghi2014}. In Ref.~\citenum{Lee2011} the observed effect of the carbon stemming from the organic solvents is pointed out. It can form an amorphous layer between the CIGS layer and the Mo back contact, thus adding to the series resistance of the circuit. Moreover, residual carbon can be present in the entire film and is observed to limit the crystal growth. The authors minimize the presence of these carbon impurities by means of a three-step annealing process.\\
\indent It is however not clear from the experimental studies what influence carbon-related point defects exert on the electric properties of CIGS. We have previously studied from first-principles which native point defects contribute to the conductivity in CIGS \cite{Bekaert2014}. Here, we have shown that the cation-related point defects, namely the vacancy V$_{\mathrm{Cu}}$, the vacancy V$_{\mathrm{In}/\mathrm{Ga}}$ and the antisite defect Cu$_{\mathrm{In}/\mathrm{Ga}}$ act as shallow acceptors, very likely giving rise to the three acceptor levels observed with photoluminescence, e.g.~in Ref.~\citenum{Siebentritt13}. On the other hand, the shallow donor In/Ga$_{\mathrm{Cu}}$ is also abundantly present in samples grown under In/Ga-rich conditions. This donor compensates to a large extent the acceptor defects, resulting in potential fluctuations through the material, also observed in photoluminescence spectra \cite{Siebentritt13}. Throughout this study the hybrid HSE functional was used for a better account of the band gap compared with standard density functional theory (DFT) \cite{Heyd}. The hybrid functional has not only been applied to CIGS in our study but also by L.~E.~Oikkonen et al.~\cite{Oikkonen2012} and J.~Pohl et al.~\cite{Pohl2013}. Yet, to the best of our knowledge no first-principles study discussing the role of carbon impurities in CIGS is available in the scientific literature. We present calculations of several types of point defects in the limiting compounds CuInSe$_{2}$ (CIS) and CuGaSe$_{2}$ (CGS). More specifically, we discuss the structural and electronic properties of the substitutional defect C$_{\mathrm{Cu}}$ in CIS and CGS, the substitutional defects C$_{\mathrm{In}}$ and C$_{\mathrm{Ga}}$ in CIS and CGS respectively and the C interstitial (C$_{\mathrm{i}}$) in both materials. The possibility to form a C interstitial can be understood in terms of the atomic radii. The C atom has a radius of $\sim 0.77$ \AA~(defined in the tetrahedral covalent bond) \cite{Kittel}, which is small compared with the interatomic distances in CIS and CGS. To start with, we review the important concepts of a first-principles study of defects, such as the formation energy, transition levels etc. Subsequently, we describe how the lattice is deformed by the C impurities and proceed by studying their electronic activity based on the formation energy of different possible charge states. In this way we ultimately try to answer whether C-related defects easily form in CIGS and what effect they have on the performance of CIGS as a photovoltaic absorber material.

\section{Theoretical method to study C-related point defects}

\subsection{Formation energy, defect concentration}

In general, the formation energy of a defect $\mathcal{D}$ in charge state $q$, $E_f\left(\mathcal{D},q\right)$ - as can be found in many publications including \cite{Bekaert2014,Ma2011,Amini} - is defined as
\begin{align}
E_f\left(\mathcal{D},q\right)=E_{\mathrm{tot}}\left(\mathcal{D},q\right)-E_{\mathrm{tot}}\left(\mathrm{bulk}\right)+\sum_{\nu}n_{\nu}\mu_{\nu} 
\notag \\ 
+q\left(E_{VBM}+E_F+\Delta V^{(q)}\right)~.
\label{eq:eformation}
\end{align}
In this expression, $E_{\mathrm{tot}}\left(\mathcal{D},q\right)$ is the total energy of a supercell containing the defect and $E_{\mathrm{tot}}\left(\mathrm{bulk}\right)$ is the total energy of the bulk supercell (i.e.~without defect). In the third term, $\mu_{\nu}$ are the chemical potentials of the atoms that are exchanged with external reservoirs to form the defect. The absolute value $\left|n_{\nu}\right|$ give the number of exchanged atoms of element $\nu$; furthermore if the atoms are added to the system $n_{\nu}<0$, in case they are removed $n_{\nu}>0$. For instance, for the C$_{\mathrm{Cu}}$ antisite defect this term is $\mu_{\mathrm{Cu}}-\mu_{\mathrm{C}}$. The chemical potentials depend as such on the chemical conditions during the growth of the material. The chemical potential can be rewritten as the sum of the chemical potential of the elemental phase ($\mu_{\nu}^{\mathrm{elem}}$) and a deviation $\Delta \mu_{\nu}$, where a more negative $\Delta \mu_{\nu}$ means $\nu$-poorer growth conditions. The range of the $\Delta \mu_{\nu}$ of the atoms making up the host material (in this case CIGS) is restricted in thermodynamic equilibrium (cfr.~Ref.~\citenum{Bekaert2014}). For charged defects, the last term in Eq.~\ref{eq:eformation} describes the exchange of electrons with the electron reservoir at the Fermi level $E_F$, referenced to $E_{VBM}$, the top of the valence band of the bulk cell ($q<0$ if they are added to the supercell and $q>0$ if they are removed). Finally, $\Delta V^{(q)}$ is the difference in reference potential of the supercell without defect and with defect. Eq.~\ref{eq:eformation} thus states that the formation energies of the defects are linear functions of $E_F$ and this is how we will represent them. The Fermi level at which the formation energies of different charge states $q$ and $q'$ of a certain defect become equal is called the transition level $\varepsilon\left(\mathcal{D},q/q'\right)$. The transition levels relative to the valence and conduction band determine the electrical activity of the defect state. The equilibrium concentration of defects of type $D$ in charge state $q$ follows a Boltzmann distribution \cite{Bekaert2014,Ma2011}:
\begin{align}
N(\mathcal{D},q)=M_{\mathcal{D}}~g_q~\mathrm{exp}\left[-E_f(\mathcal{D},q)/(k_B T)\right]~,
\label{eq:defconcentr}
\end{align}
where $M_{\mathcal{D}}$ denotes the concentration of lattice sites where the defect can originate and $g_{q}$ is a degeneracy factor for charge state $q$, dependent on the electronic degeneracy, including spin degeneracy~\cite{Bekaert2014}. For example, for C$_{\mathrm{Cu}}$ in CIS this multiplicity is the number of Cu lattice sites per cm$^{3}$, i.e.~$\sim 1.13 \cdot 10^{22}$ cm$^{-3}$.

\subsection{Computational details}

As mentioned in the Introduction, our calculations use a hybrid functional, more specifically the HSE06 functional implemented in the VASP code \cite{Kresse,Paier}. With the standard HSE06 functional the exchange interaction is still overscreened, resulting in underestimated band gaps amounting to 0.85 eV for CIS and 1.37 eV for CGS. We have determined that the experimental band gaps \cite{Belhadj} correspond to an enhanced fraction of Hartree-Fock exchange of $\alpha=0.2780$ for CIS and $\alpha=0.3098$ for CGS producing gaps of 1.00 eV and 1.72 eV respectively. We use the adapted HSE06 functional throughout this paper since correct values for the band gaps are crucial for calculating defect formation energies. Electron-ion interactions are treated using projector augmented wave (PAW) potentials, including Cu-3d$^{10}$4s$^1$, Ga-3d$^{10}$4s$^2$4p$^1$, In-4d$^{10}$5s$^2$5p$^1$, Se-4s$^2$4p$^4$ and C-2s$^{2}$2p$^2$ as valence electrons. The energy cutoff for the plane-wave basis is set to 500 eV. The C impurities are placed in a supercell of the primitive cell, to reduce electrostatic interactions between the impurities adding to the total energy. Previously, we have performed a convergence test for the supercell size finding that the $2\times 2 \times 2$ supercell of the primitive cell of CIGS, containing 64 atoms, yields well-converged values \cite{Bekaert2014}. Thus, we present formation energies related to the C-impurities calculated in 64-atom supercells here. The atomic positions in these supercells are relaxed until all forces are smaller than 0.05 eV/\AA, while keeping the cell volume fixed. The charge state $q$ of the impurity is simulated by adding $q=.., -2, -1, 0, +1, +2, ..$ electrons to the supercell. For integration over the Brillouin zone a $2 \times 2\times 2$ $\Gamma$-centered Monkhorst-Pack \textbf{k}-point grid is used. Finally, we calculate the correction for the reference potential $\Delta V^{(q)}$ in Eq.~\ref{eq:eformation} via the method described in Ref.~\citenum{Amini}. 

\section{Results and discussion}

\subsection{Structural properties}

\begin{figure}[t]
\centering
\includegraphics[width=0.23\textwidth]{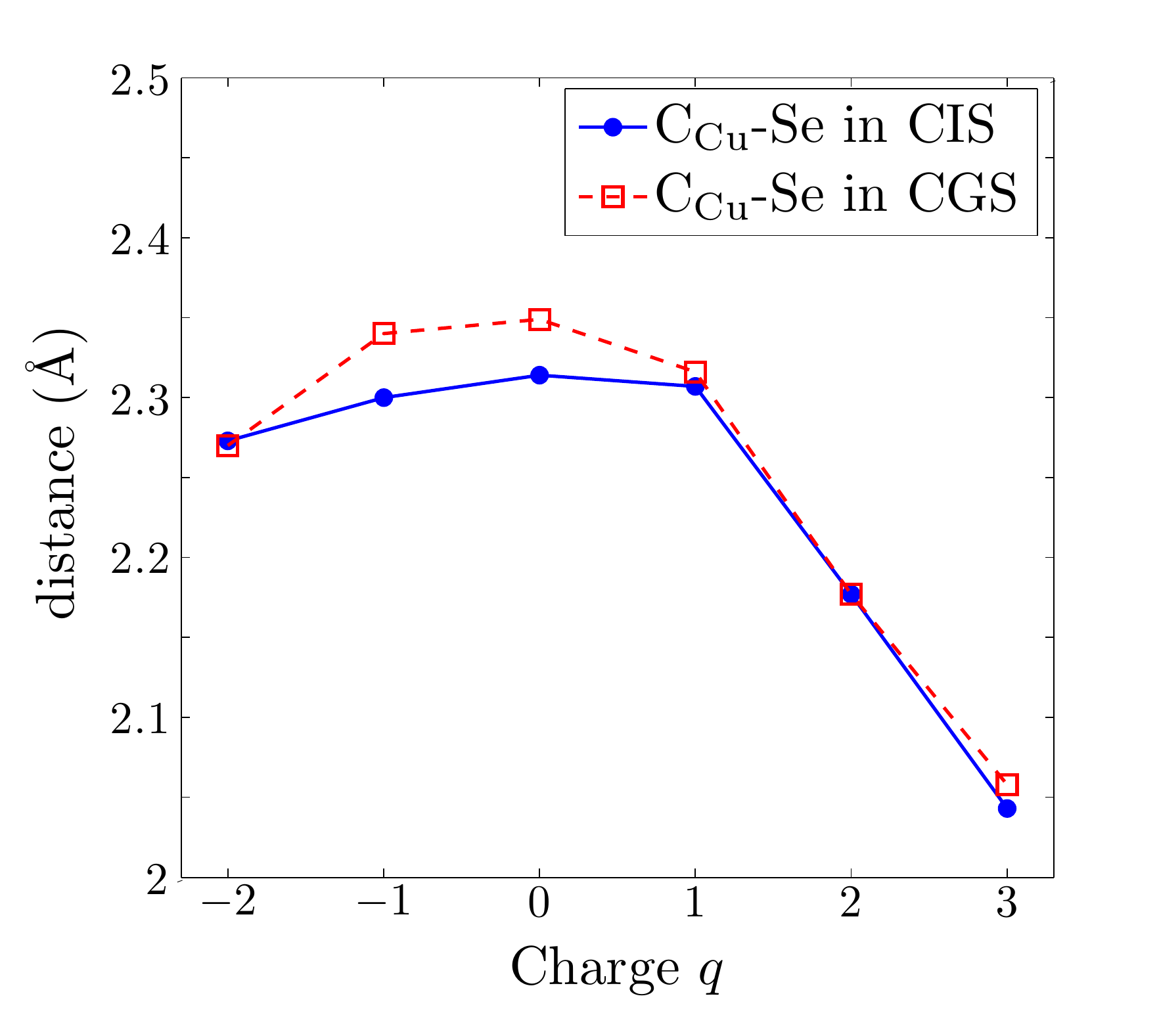}\llap{
  \parbox[b]{3.2in}{(a)\\\rule{0ex}{1.35in}
  }}
\includegraphics[width=0.22\textwidth]{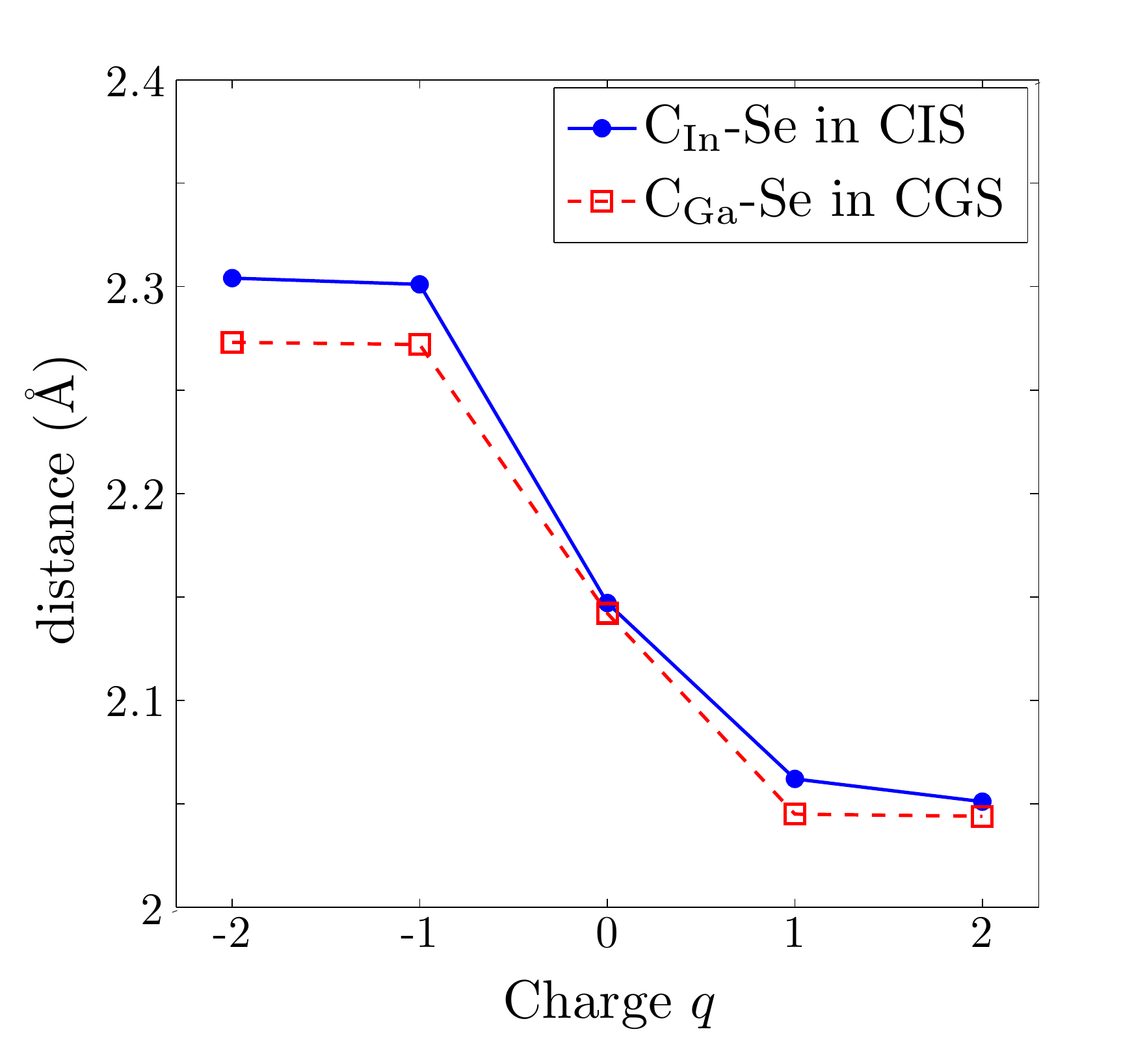}\llap{
  \parbox[b]{3.05in}{(b)\\\rule{0ex}{1.35in}
  }}
  
  \vspace{0.2 cm}
  
\includegraphics[width=0.23\textwidth]{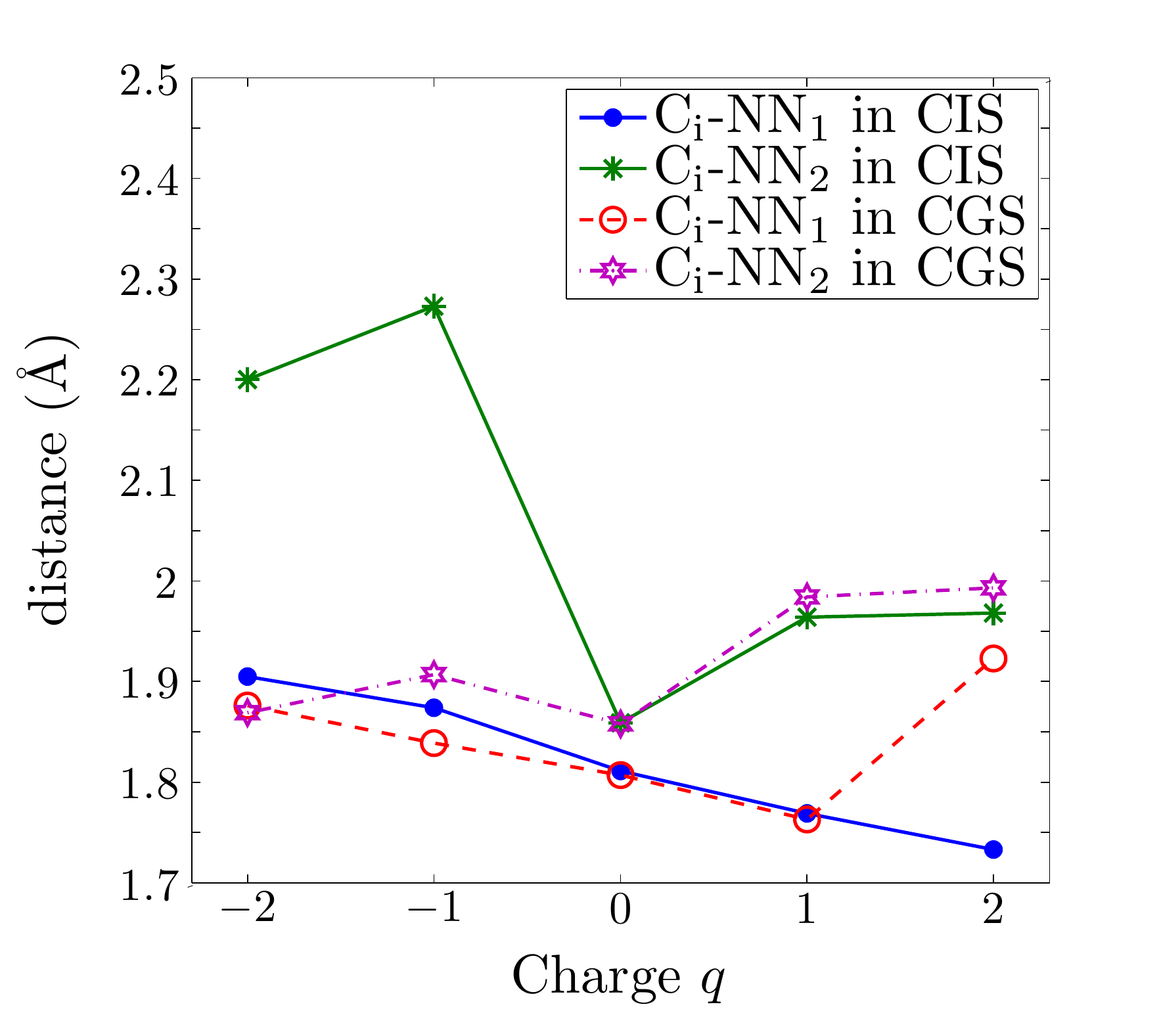}\llap{
  \parbox[b]{3.2in}{(c)\\\rule{0ex}{1.35in}
  }}
\caption{(Color online) The interatomic distances (\AA) between the C impurities and the surrounding atoms as a function of the excess charge $q$ of the impurity. The lines connecting the values for different charge states serve as a guide for the eye. In (a) the distance C$_{\mathrm{Cu}}$-Se is given, in (b) the distances C$_{\mathrm{In}}$-Se and C$_{\mathrm{Ga}}$-Se and in (c) the distance between C$_{\mathrm{i}}$ and the two nearest neighbors (NN). These are NN$_1=\mathrm{Se}$ in both CIS and CGS for all charge states and NN$_2=\mathrm{Cu}$ except for $q=-1,-2$ in CGS where the second nearest neighbor is Ga. For comparison, the calculated unperturbed interatomic distances are $d\left(\mathrm{Cu-Se}\right)=2.456$ \AA~in CIS and $d\left(\mathrm{Cu-Se}\right)=2.440$ \AA~in CGS and $d\left(\mathrm{In-Se}\right)=2.609$ \AA~in CIS and $d\left(\mathrm{Ga-Se}\right)=2.429$ \AA~in CGS.}
\label{fig:distances}
\end{figure}
\begin{figure}[t]
\centering
\includegraphics[width=0.32\textwidth]{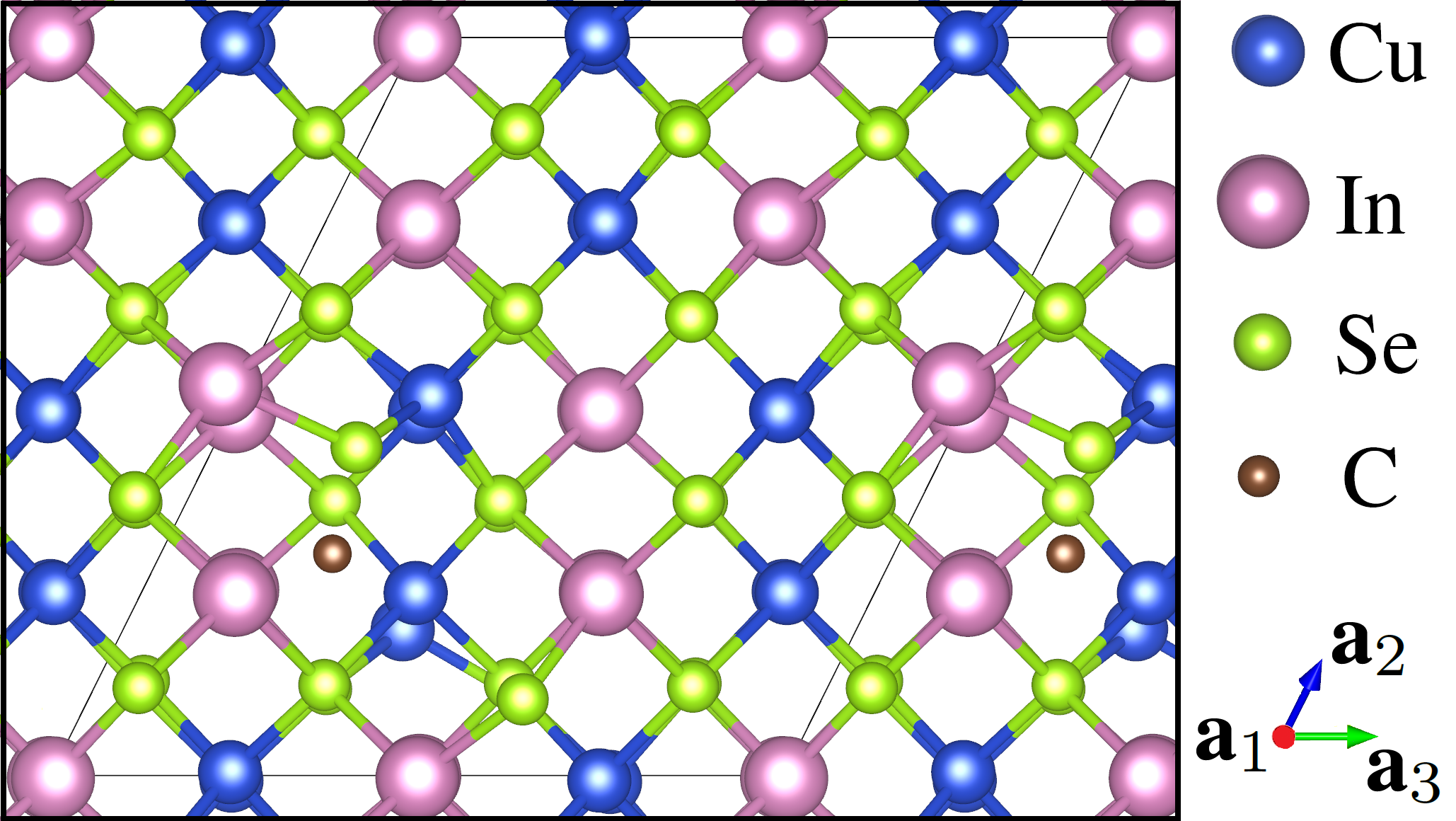}
\caption{(Color online) The lattice distortion in CIS due to the C$_{\mathrm{i}}$ defect. The $2\times2\times2$ supercell is indicated by the gray lines. The position of the interstitial in terms of $\textbf{a}_1=(2a,0,0)$, $\textbf{a}_2=(0,2a,0)$ and $\textbf{a}_3=(a,a,c)$, the primitive lattice vectors that span the $2\times2\times2$ supercell, is $\sim (0.97,0.23,0.30)$.}
\label{fig:carbon_int}
\end{figure}

The calculated lattice parameters of pristine CIS are $a=5.832$ \AA, $c=11.735$ \AA~and anion displacement $u=0.229$ and those of CGS are $a=5.652$ \AA, $c=11.119$ \AA~and $u=0.253$ \cite{Bekaert2014}. As we have already mentioned, the cell shape and volume are kept fixed after introducing the impurity. The ionic positions are relaxed and, naturally, the largest differences appear in the interatomic distances of the C impurities and the surrounding atoms. In Fig.~\ref{fig:distances} we present these distances as a function of the excess charge of the impurity, $q$. In Fig.~\ref{fig:distances} (a) one can observe that the interatomic C$_{\mathrm{Cu}}$-Se distance reaches a maximum for the neutral cell ($q=0$), but is for all charge states smaller than the unperturbed Cu-Se distance. These observations hold in both CIS and CGS. The distances C$_{\mathrm{In}}$-Se and C$_{\mathrm{Ga}}$-Se, on the other hand, converge to maximal values of $\sim 2.30$ \AA~in CIS and $\sim 2.27$ \AA~in CGS for negative $q$. Similarly, for positive $q$ the distances converge to minimal values of $\sim 2.05$ \AA~in CIS and CGS. Finally, the initial position of the C$_{\mathrm{i}}$ defect is chosen such that the distances to all other atoms in the lattice is maximal. An example of the structure after relaxation is shown in Fig.~\ref{fig:carbon_int} for CIS in the $q=0$ charge state. In fact, for $q=0$ the distances between C$_{\mathrm{i}}$ and the first nearest neighbor NN$_1=\mathrm{Se}$ are almost equal in CIS and CGS; this also holds for the second nearest neighbor NN$_2=\mathrm{Cu}$. The distances to C$_{\mathrm{i}}$-Se are overall linearly decreasing with increasing charge state, with the exception of $q=-2$ in host CGS, where there is an increase of this interatomic distance of $0.16$ \AA. The distance C$_{\mathrm{i}}$-NN$_2$ (NN$_2=\mathrm{Cu}$ except for $q<0$ in host CGS, where NN$_2=\mathrm{Ga}$) reaches a minimum for $q=0$. For positive charge states this interatomic distance converges to $\sim 2$ \AA~in CIS and CGS. For the negative charge states, however, there is a strong increase to values exceeding $2.2$ \AA~in CIS, whereas the C$_{\mathrm{i}}$-Ga distance remains around $1.9$ \AA~in CGS. In general, one can summarize that the lattice deformation due the substitutional C impurities are similar in CIS and CGS for all charge states, but not very surprisingly the lattice deformations due to the interstitial C are more complex. This complex behavior seems to be dominated by the excess charge states, since in the neutral case the interatomic distances are almost equal in CIS and CGS. 

\subsection{Electronic properties}

\begin{figure}[t]
\centering
\includegraphics[width=0.24\textwidth]{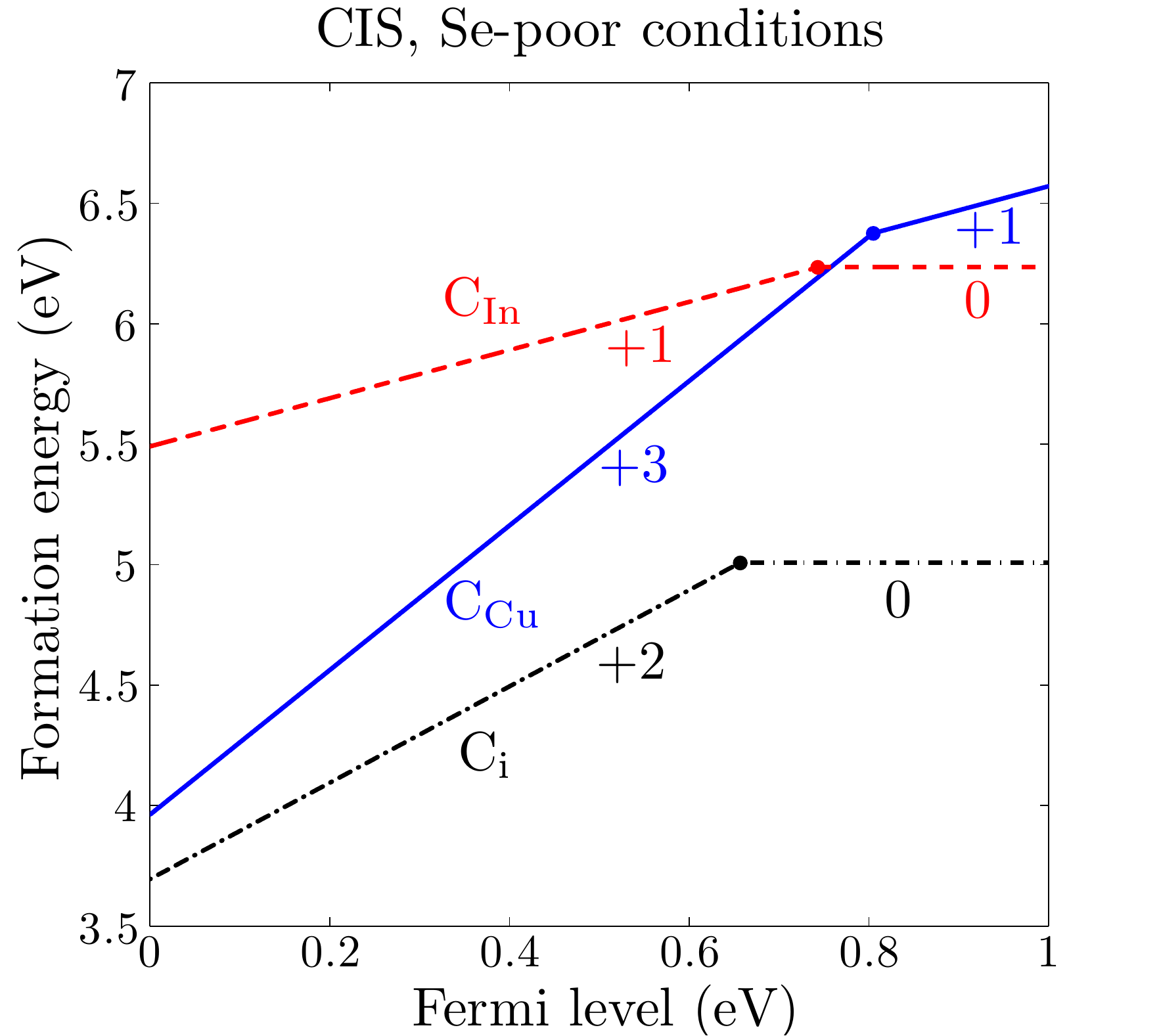}\llap{
  \parbox[b]{3.3in}{(a)\\\rule{0ex}{1.5in}
  }}
\includegraphics[width=0.235\textwidth]{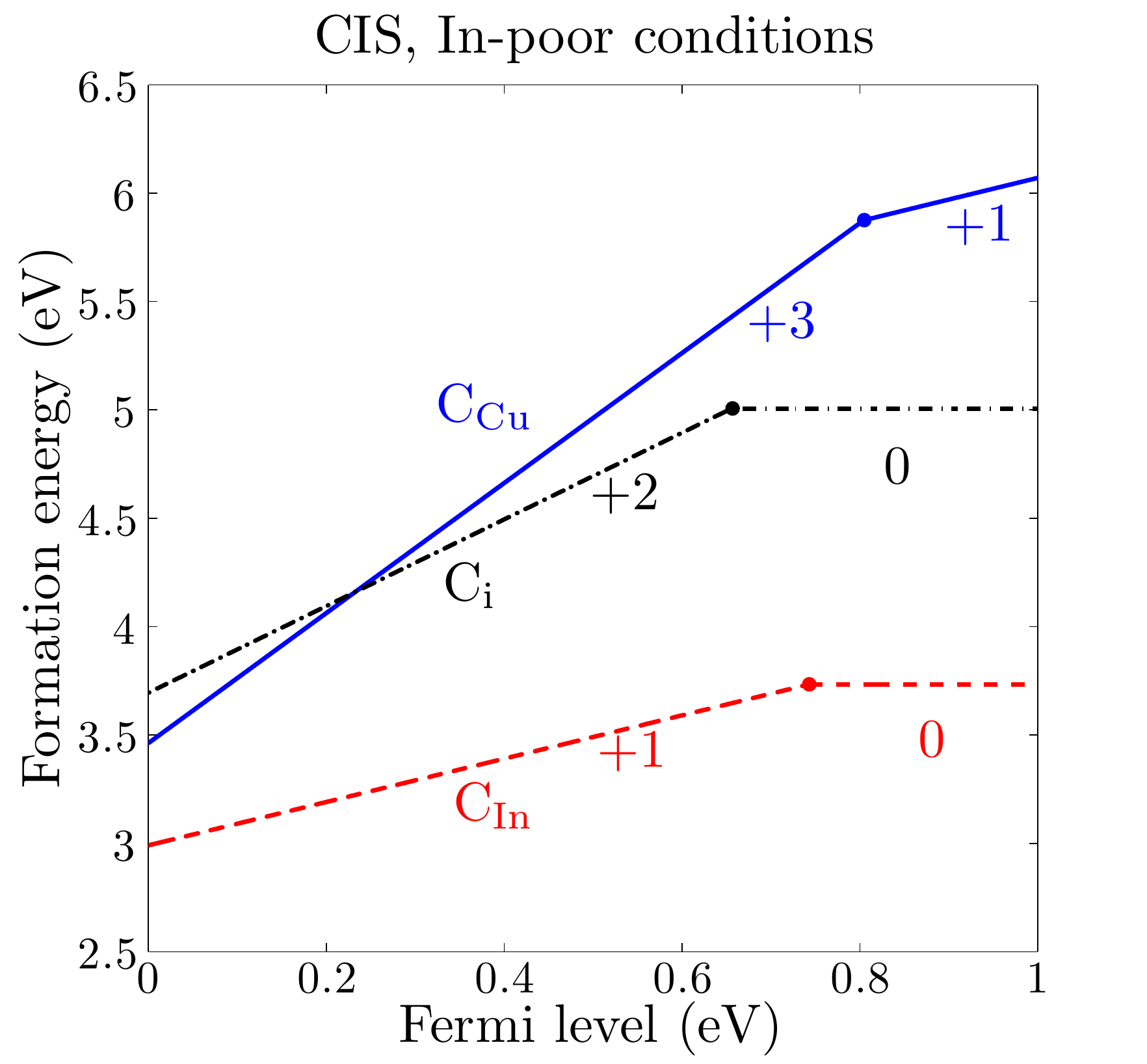}\llap{
  \parbox[b]{3.3in}{(b)\\\rule{0ex}{1.5in}
  }}

\vspace{0.35 cm}

\includegraphics[width=0.24\textwidth]{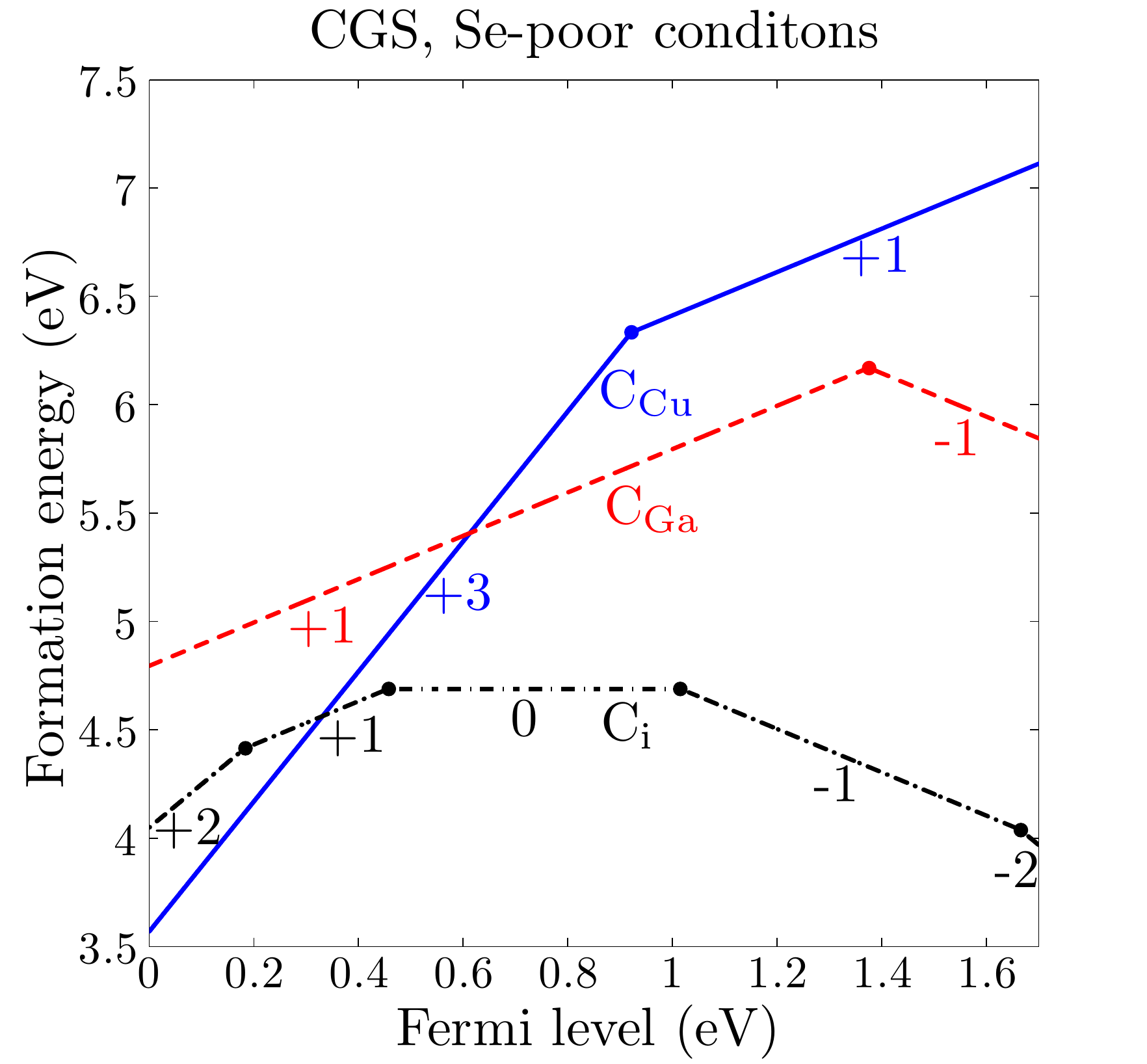}\llap{
  \parbox[b]{3.3in}{(c)\\\rule{0ex}{1.5in}
  }}
\includegraphics[width=0.235\textwidth]{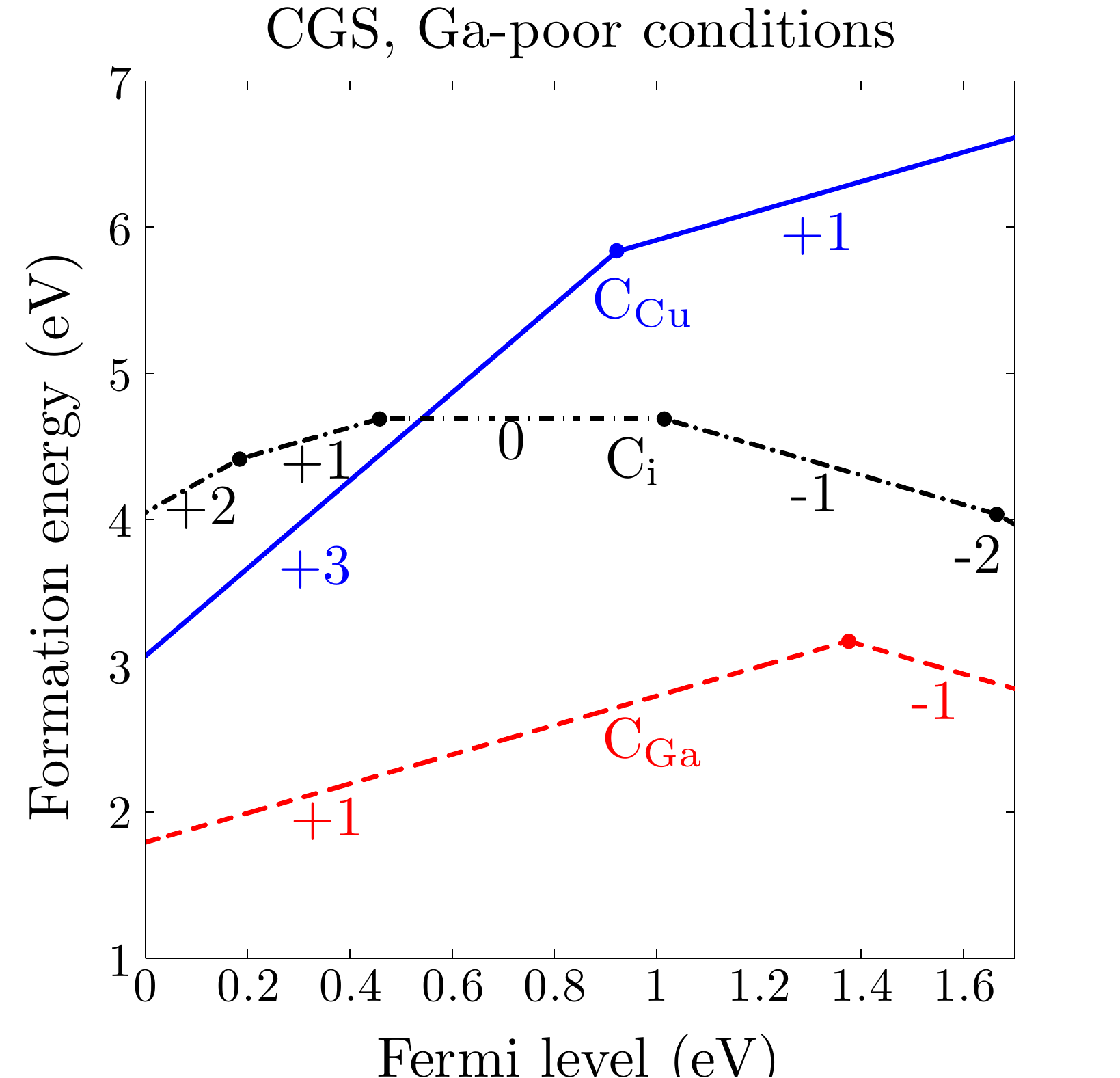}\llap{
  \parbox[b]{3.3in}{(d)\\\rule{0ex}{1.5in}
  }}
\caption{(Color online) Formation energies (eV) of ground charge states of C$_{\mathrm{\mathrm{Cu}}}$ (idem), C$_{\mathrm{In}}$ (C$_{\mathrm{Ga}}$) and C$_{\mathrm{i}}$ (idem) in CIS (CGS), as function of the Fermi level between VBM and CBM. The charge states are listed near the curves, while the transition levels are indicated by solid dots. For CIS, we distinguish between (a) Se-poor conditions $\left(\Delta \mu_{\mathrm{Cu}},\Delta \mu_{\mathrm{In}}\right)=\left(0,0\right)$ eV and (b) In-poor conditions $\left(\Delta \mu_{\mathrm{Cu}},\Delta \mu_{\mathrm{In}}\right)=\left(-0.5,-2.5\right)$ eV. Similarly, for CGS we compare (c) Se-poor conditions $\left(\Delta \mu_{\mathrm{Cu}},\Delta \mu_{\mathrm{Ga}}\right)=\left(0,0\right)$ eV with (d) Ga-poor conditions $\left(\Delta \mu_{\mathrm{Cu}},\Delta \mu_{\mathrm{Ga}}\right)=\left(-0.5,-3.0\right)$ eV. In all cases $ \mu_{\mathrm{C}}^{\mathrm{graphite}}$ is used to calculate the exchange of C atoms.}
\label{fig:form_en}
\end{figure}

The formation energies of the ground charge states of the different C impurities are plotted as a function of the Fermi level $E_F$ in Fig.~\ref{fig:form_en}, under different chemical growth conditions via the chemical potentials of the exchanged atoms. The transitions between the charge states are however not affected by the chemical potentials, as follows from Eq.~\ref{eq:eformation}. The formation energies of C$_{\mathrm{Cu}}$ show that it prefers to donate electrons. For small $E_F$, the $q=+3$ charge state is favored owing to the 3 extra outer shell electrons of C compared with Cu. For higher $E_F$ donating 3 electrons becomes too energetically expensive, resulting in a direct transition to the $q=+1$ charge state at 0.81 eV for CIS and a similar transition at 0.92 eV in CGS. These transition levels are also listed in Table \ref{tab:trans_levels}. Since the transition to the neutral charge states lies within the conduction band of the host material, C$_{\mathrm{Cu}}$ acts as a shallow donor in both CIS and CGS. Moreover, it can compensate the native acceptor-type defects in p-type CIGS, thereby reducing the free hole concentration $p$. The substitutional defects C$_{\mathrm{In}}$ and C$_{\mathrm{Ga}}$ adopt the $q=+1$ in the lower range of $E_F$. The interpretation is again clear: by donating one electron, C carries the same number of outer shell electrons as In and Ga. Also, with increasing $E_F$ donating electrons becomes energetically unfavorable and C$_{\mathrm{In}}$ undergoes a +1/0 transition at $E_F=0.74$ eV (0.26 eV from the CBM) in CIS. In CGS, there is a direct +1/-1 transition at $E_F=1.38$ eV (0.34 eV from the CBM). Since C$_{\mathrm{Ga}}$ acts as both a deep acceptor and a deep donor, it is also called a deep amphoteric defect \cite{Freysoldt14}. So, in p-type CIS and CGS, C$_{\mathrm{In}}$ and C$_{\mathrm{Ga}}$ give rise to deep donor levels, that compensate acceptor levels. As a result, they become positively charged and subsequently act as traps for the photogenerated minority carriers (electrons). C$_{\mathrm{i}}$ in CIS donates two electrons provided that $E_F<\varepsilon\left(+2/0\right)=0.66$ eV (0.34 eV from the CBM). This means that C$_{\mathrm{i}}$ acts as a deep donor in CIS. In CGS, C$_{\mathrm{i}}$ also acts donor-like for the lower range of $E_F$. There is a transition from $q=+2$ to $q=+1$ at 0.18 eV. Subsequently, at 0.46 eV the neutral state becomes the ground state. For the higher range of $E_F$ in the gap the defect favors acceptor behavior. The transition levels are $\varepsilon\left(0/-1\right)=1.02$ eV and $\varepsilon\left(-1/-2\right)=1.67$ eV and therefore C$_{\mathrm{i}}$ acts as a deep amphoteric defect in CGS. It is thus a deep level trap for either charge carrier type, depending on $E_F$. In p-type CIS and CGS, C$_{\mathrm{i}}$ can again compensate acceptor levels, thereby becoming a trap for photogenerated electrons. In summary, C impurities could be detrimental for the performance of p-type CIS and CGS in a photovoltaic device, since they act as shallow donors (C$_{\mathrm{Cu}}$) or deep donors (C$_{\mathrm{In}}$, C$_{\mathrm{Ga}}$ and C$_{\mathrm{i}}$). They compensate acceptor levels and become trap states for the photogenerated minority carriers.\\
\begin{table}[t]%
\caption{\label{tab:trans_levels}%
Transition levels (eV) within the band gap, given w.r.t.~to the VBM.}
\begin{ruledtabular}
\begin{tabular}{cclc}
\textrm{Host}&
\textrm{Defect}&
\multicolumn{1}{c}{\textrm{Transition $q/q'$}}&
\textrm{$\varepsilon\left(q/q'\right)$ (eV)}\\
\colrule
CIS&C$_{\mathrm{Cu}}$&$+3/+1$&0.81\\ 
CGS&C$_{\mathrm{Cu}}$&$+3/+1$&0.92\\ 
CIS&C$_{\mathrm{In}}$&$+1/0$&0.74\\ 
CGS&C$_{\mathrm{Ga}}$&$+1/-1$&1.38\\ 
CIS&C$_{\mathrm{i}}$&$+2/0$&0.66\\ 
CGS&C$_{\mathrm{i}}$&$+2/+1$&0.18\\ 
CGS&C$_{\mathrm{i}}$&$+1/0$&0.46\\ 
CGS&C$_{\mathrm{i}}$&$0/-1$&1.02\\ 
CGS&C$_{\mathrm{i}}$&$-1/-2$&1.67\\ 
\end{tabular}
\end{ruledtabular}
\end{table}
\indent There is however an important additional consideration we have to make, namely the number of C impurities that form depends on the formation energy according to Eq.~\ref{eq:defconcentr}. The values of the formation energy depend on the chemical growth conditions via the chemical potentials. In Fig.~\ref{fig:form_en} (a) the formation energies are calculated under Se-poor conditions. In this case, $E_F$ is pinned close to the CBM because the donor In$_{\mathrm{Cu}}$ has a low formation energy \cite{Bekaert2014}. Under these conditions, the defect with the lowest formation energy is C$_{\mathrm{i}}$, but its formation energy still amounts to $\sim$ 5 eV in n-type material CIS. Moreover, it is not electrically active for $0.66~\mathrm{eV}<E_F<1.00$ eV. On the other hand, to obtain p-type conductivity based on native point defects, more In-poor conditions are required, such as those leading to Fig.~\ref{fig:form_en} (b). Owing to the In-poor conditions C$_{\mathrm{In}}$ has become the dominant C impurity. However, the formation energy of this defect is $\sim 3$ eV if $E_F$ is close to the VBM. Using Eq.~\ref{eq:defconcentr} this leads to a negligible concentration of C$_{\mathrm{In}}$. In Fig.~\ref{fig:form_en} (c) we display the formation energies of the C impurities in CGS grown under Se-poor conditions. Like in CIS, C$_{\mathrm{i}}$ is the defect with the lowest formation energy under these conditions. However, the formation energies are very high, exceeding 3.5 eV, so the impurities are very unlikely to form. Under Ga-poor conditions the native defects pin $E_F$ at the VBM, so based on the formation energies in Fig.~\ref{fig:form_en} (d) C$_{\mathrm{Ga}}$ is the dominant C impurity. Still, its formation energy is $\sim$ 1.8 eV, corresponding to less than one C$_{\mathrm{Ga}}$ defect per cm$^{3}$. So, in all possible chemical growth conditions for CIGS the formation energy of C impurities is very high, so the resulting impurity concentration is negligible.\\
\indent In the results presented in Fig.~\ref{fig:form_en} we have assumed that the reservoir for C atoms is the elemental solid, graphite. One may wonder how the results are affected by a different choice of reservoir. In reality, the C atoms are part of the organic molecules of the solution used to disperse the nanoparticle precursor. Methanol (CH$_3$OH) is probably the most simple organic solvent used in these solutions \cite{Schulz1998}. It can readily be studied using the hybrid functional, enclosing it in a box, i.e.~a supercell with an edge of 30 $\mathrm{\AA}$ that is otherwise empty. In order to calculate the formation energy of CH$_3$OH, the total energy of graphite is to be taken into account. The total energies of the O$_2$ and H$_2$ molecules have to be included as well. For this computation, O$_2$ and H$_2$ are enclosed in boxes with edges measuring 30 $\mathrm{\AA}$. The total energy of O$_2$ is obtained in a spin-polarized calculation, since the triplet state with two unpaired electrons is the ground state. The calculated value of the formation energy of methanol is thus
\begin{align}
\Delta H_f(\mathrm{CH}_3\mathrm{OH})&=E_{tot}(\mathrm{CH}_3\mathrm{OH})-\frac{E_{tot}(\mathrm{graphite})}{\mathrm{atom}} \notag \\  & \hspace{4mm} -\frac{1}{2}E_{tot}(\mathrm{O}_2)-2E_{tot}(\mathrm{H}_2) \notag \\
&=-2.71~\mathrm{eV}~.
\end{align}
An experimentally obtained value of the heat of formation in the liquid phase is $-238.4$ kJ/mol \cite{NIST}. This corresponds to $-2.47$ eV per CH$_3$OH molecule, yielding a good agreement with the theoretical value. In order that the C atoms of CH$_3$OH do not precipitate into graphite one requires that $\Delta \mu_{\mathrm{C}} \leq 0$. Consequently, the chemical potential $ \mu_{\mathrm{C}}$ ranges from $ \mu_{\mathrm{C}}^{\mathrm{graphite}}$ under C-rich conditions to $ \mu_{\mathrm{C}}^{\mathrm{graphite}}-2.71$ eV under C-poor conditions. We find that the formation energies can shift upwards with $+2.71$ eV at most (under C-poor conditions, which are not plausible). Thus, we can conclude that a reservoir containing organic molecules (an analogous reasoning can be applied to other solvents) leads to an additional increase of the formation energy of C impurities.\\
\indent This leads to the overall conclusion that the formation of C impurities in CIGS is very unlikely. Hence, it can be expected that C is expelled outside of CIGS, to the grain boundaries or outside of the absorber layer. It corroborates the experimental observation of a thick amorphous C layer forming between the CIGS layer and the substrate \cite{Lee2011}. On the other hand, we are not aware of a comprehensive experimental study of C impurities at the grain boundaries of CIGS.

\section{Conclusion}

We have studied C impurities in CIGS, related to new nonvacuum growth methods using e.g.~nanoparticle inks. We have calculated the formation energies of several substitutional impurities and also an interstitial impurity using the hybrid functional HSE06. We found that C$_{\mathrm{Cu}}$ acts as a shallow donor in both CIS and CGS. C$_{\mathrm{In}}$ and interstitial C yield deep donor levels in CIS, while in CGS C$_{\mathrm{Ga}}$ and interstitial C act as deep amphoteric defects. Therefore, in p-type CIS and CGS these impurities can compensate acceptor levels, thus reducing the majority carrier concentration, and become trap states for the minority carriers. As such, they are in principle harmful to the performance of CIGS photovoltaic device. However, we observe that the formation energy of C defects is very high, even under C-rich conditions. Consequently, C defects are not likely to be formed in CIGS. We expect that C is expelled out of the CIGS grains in nonvacuum growth methods, to the grain boundaries and outside of the absorber layer.

\section*{Acknowledgement}

We gratefully acknowledge financial support from the science fund FWO-Flanders through project G.0150.13. J.~B.~would furthermore like to thank the University of Antwerp for the grant he receives for his PhD study. The first-principles calculations have been carried out on the HPC infrastructure of the University of Antwerp (CalcUA), a division of the Flemish Supercomputer Centre (VSC), supported financially by the Hercules foundation and the Flemish Government (EWI Department). 

\bibliography{biblio}

\end{document}